\documentclass[preprint, review, 12pt]{elsarticle}

\usepackage{amssymb}
\usepackage{amsmath,bm}
\usepackage{colortbl,booktabs}
\usepackage{tabularx}
\usepackage{threeparttable}
\usepackage{multicol,multirow}
\usepackage{subfigure}
\usepackage [font=footnotesize,labelfont=bf]{caption}
\usepackage{rotating}
\usepackage{tikz}
\usepackage{hyperref}

\journal{Elsevier}

\begin{document}
	\begin{frontmatter}
		
		\title{Heat transfer simulation of window frames with SPHinXsys}
		
		\author{Haotian Ji \fnref{eqnote}}
		\ead{haotian.ji@tum.de}
		\author{Dong Wu \fnref{eqnote}}
		\ead{dong.wu@tum.de}
		\author{Chi Zhang \fnref{eqnote}}
		\ead{c.zhang@tum.de}
		\author{Xiangyu Hu \fnref{eqnote}}
		\ead{xiangyu.hu@tum.de}
		\address{School of Engineering and Design, Technical University of Munich\\
			85748 Garching, Germany}
		
		\begin{abstract}
			Maintaining a comfortable temperature inside a building requires appropriate thermal insulation of windows, which can be optimised iteratively with numerical simulation. Smoothed particle hydrodynamics(SPH) is a fully Lagrangian method widely used for simulating multi-physics applications with high computational efficiency and accuracy. It is advantageous in physically coupled problems such as heat-fluid-solid or any other type of physically coupled simulations. The focus of this study is to simulate the heat transfer process in various window frames under convective boundary conditions according to ISO10077-2:2012. This paper demonstrates the accuracy and compatibility of SPH when dealing with heat transfer problems, which ensures further development of thermal coupling with other physical fields. The results and methods used in this paper provide some guidance on how to properly handle heat transfer simulations using SPH, which can be extended to multi-physics coupled simulations in the future.

		\end{abstract}
		
		\begin{keyword}
			Smoothed Particle Hydrodynamics \sep Heat conduction \sep 
			Thermal convection \sep Multi-material structure
		\end{keyword}
		
	\end{frontmatter}
	
	\section{Introduction}
	Heat transfer between buildings and their surroundings has a direct impact on the perception of a comfortable temperature for the people in the building. Proper selection of window frame constructions and materials, such as thermal bridges or low thermal conductivity materials, can minimise the heat losses and achieve environmental friendly features. Numerical simulation can provide effective guidance for the iterative optimisation of window frame design strategies and design solutions. The international standard ISO10077-2:2012 deals with the thermal properties of windows, doors and shutters. It provides reference values for thermal properties of window frame profiles for validating the results of numerical methods. COMSOL Multiphysics, a widely known multi-physics software, passes the entire window frame profile benchmarks illustrated in ISO10077-2:2012 and provides some guidelines for dealing with heat transfer simulation in window frames \cite{Comsol}.\\
	The most common numerical methods used for thermal simulation to solve technical problems are meshing methods such as Finite Element Method(FEM), Finite Difference Method(FDM), Finite Volume Method(FVM) and Boundary Element Method (BEM). These numerical methods have been successfully validated and implemented in many commercial software packages \cite{Mackerle2002}. Cen et al. \cite{Cen2017} proposed a numerical method based on FVM to simulate heat transfer in three-dimensional non-homogeneous materials with high accuracy.  Haseli and Naterer \cite{Haseli2022} proposed a new semi-analytical solution for transient heat transfer in slabs, where heat is exchanged with the surroundings by convection and radiation, which provides some analytical results for convection problems in thermal simulation and some guidance for numerical validation.\\
	In addition to heat transfer algorithms, there is a large body of research on the thermal performance of window systems. Werner-Juszczuk and Rynkowski \cite{Anna2016} have carried out thermal simulations of 2D heat transfer in complex multi-regions using BEM and validated the numerical algorithm according to ISO10211:2007. Onatayo et al. \cite{Onatayo2024} conducted a data driven approach to acquire thermal transmittance (U-factor) of double-glazed windows with or without inert gases between the panes. Pittie et al. \cite{Pittie2024} presented a robust and cost-effective numerical framework for designing aluminium window profiles and validated the model against experimental results. Muhic et al. \cite{Muhic2024} investigated the effect of building energy model parameter input values on simulation results to understand how building thermal envelope model parameters affect building energy consumptions.\\
	To the best of our knowledge, there are no engineering applications using SPH method to study heat transfer in complex window frames. SPH is a completely Lagrangian mesh-free method that was originally developed for solving astrophysical problems. In SPH, the continuum is represented by particles with physical properties and the discretisation of the governing equations is achieved by inter-particle interactions with the help of Gaussian kernel functions \cite{Wu2023}. There are many SPH algorithms related to heat transfer problems have been developed. By decomposing the second-order partial differential equation (PDE) into two first-order PDEs, Jeong et al. \cite{Jeong2003} modified the numerical steps involved in the SPH to take into account the implementation of thermal boundary conditions, which facilitates the treatment of multiphase systems. In order to deal with anisotropic thermal diffusion models for non-homogeneous materials, some new diffusion algorithms using arithmetic or harmonic mean conductivity have been developed to ensure the thermal stability of the computational process and to avoid unphysical temperature jumps \cite{Cleary1999} \cite{Biriukov2019}. For thermal boundary conditions, Fraser et al. \cite{Fraser2016} described a robust and efficient adaptive thermal boundary algorithm that considers different boundary types such as Dirichlet (for a given temperature), Neumann (for a given heat flux), and Robin (convection). In addition to SPH algorithms for heat transfer problems, there are also many engineering heat transfer applications based on SPH. Afrasiabi et al. \cite{Afrasiabi2021} developed a higher order SPH thermal model for metal cutting. Li et al. \cite{Li2021} provided a symmetric SPH method for modelling transient heat transfer in functionally graded structures. Tang et al. \cite{Tang2022} demonstrated a two-stage Runge-Kutta integrator scheme for the study of thermal-fluid-structure interaction (TFSI) problems.\\
	One novelty of this paper is that we can easily define multi-material structures for the heat transfer simulation with convenient SPH algorithms. In current engineering simulations, it has always been a challenge to define a large number of material properties for complex structures. In addition, numerical stability is also a problem brought by multi-material definition. In this paper, we can solve this problem effectively while ensuring numerical stability and simulation accuracy.\\
	For the general heat transfer process of complex window frames presented in this paper, we performed the overall simulation using an open-source SPH multi-physics library called SPHinXsys. Algorithms for handling multiple material definition, applying convective boundary condition with air condensation, and calculating equivalent thermal conductivities and steady state heat fluxes have been developed, with which we have successfully demonstrated the robustness and trustworthiness of SPH method in simulating heat transfer engineering problems and have provided numerical guidance for other types of SPH heat transfer simulations. On the basis of this paper, there is a great potential for future research in thermal coupling multi-physics modelling using SPH.
	\section{Governing equation}
	The governing equation of heat transfer for solids is often referred to as the heat diffusion equation, which is a second order partial differential equation \cite{Biriukov2019}
	\begin{equation}
		\frac{du}{dt} = \frac{1}{\rho} \bigtriangledown \cdot (k \bigtriangledown T), \label{eq:1}
	\end{equation}
	where $\rho$ is the density, $T$ is the temperature, and $k$ is the thermal conduction coefficient,
	\begin{equation}
		\frac{du}{dt} = c_{v}\frac{dT}{dt}, \label{eq:add1}
	\end{equation}
	is the time derivative of internal energy per unit mass, $c_{v}$ is volumetric heat capacity.
	We can reformulate the above equation as follows to compute diffusive flux explicitly
	\begin{equation}
		\begin{cases}
		\rho c_{v} \frac{dT}{dt} = -\bigtriangledown q \\
		q = -k \bigtriangledown T
		\end{cases}.
	\end{equation}
	In our case, the boundary condition for all window frames is convection, the heat generated by convective boundary condition is
	\begin{equation}
		-k \frac{dT}{dx}|_{x=0} = h (T_{\infty} - T_{s}),
	\end{equation}
	where $T_{\infty}$ is temperature far away from the wall, $T_{s}$ the temperature on the wall surface, and $h$ the heat convection coefficient. Then the flux form of heat diffusion equation can be rewritten into a more specific form
	\begin{equation}
		\begin{cases}
		\rho c_{v} \frac{dT}{dt} = -\bigtriangledown q \\
		q = -k \bigtriangledown T + h (T_{\infty} - T_{s})
		\end{cases}.
	\end{equation}
	The heat flux $q$ now consists of internal heat conduction and heat convection at the walls. Taking these two equations together, and considering that for convenience both $\rho$ and $c_{v}$ are equal to 1 (density and heat capacity do not affect the steady state temperature field), we can unify the above equation as
	\begin{equation}
		\frac{dT}{dt} = \bigtriangledown \cdot (k \bigtriangledown T) - \bigtriangledown \cdot (h (T_{\infty} - T_{s})). \label{eq:7}
	\end{equation}
	\section{Methodology}
	\subsection{Fundamentals of SPH}
	At the core of SPH is an interpolation method that allows any function to be represented by the values of an unordered set of particles \cite{Monaghan1992}.
	The integral interpolation of any function $A(r)$ is defined by 
	\begin{equation}
		A(r) = \int A(r') W(r-r', h) dr'.
	\end{equation}
	Then the interpolated value of a function $A$ at position $r$ is given by
	\begin{equation}
		A(r) = \sum m_{b} \frac{A_{b}}{\rho_{b}} W(r-r_{b},h).
	\end{equation}
	The gradient of the function $A(r)$ can be obtained by analytically differentiating the interpolation formula as
	\begin{equation}
		\bigtriangledown A(r) = \sum _{b} m_{b} \frac{A_{b}}{\rho_{b}} \bigtriangledown W(r-r_{b},h).
	\end{equation}
	Considering accuracy and stability, an appropriate SPH approximation of Laplace operator can be written as \cite{Fraser2016} \cite{Bonet2002}
	\begin{equation}
		\bigtriangleup A(r) = 2 \sum ^{N_{i}} _{j=1} \frac{m_{j}}{\rho_{j}} \frac{A_{i} - A_{j}}{|r_{ij}|^{2}} r_{ij} \frac{dW_{ij}}{dr_{i}^{\beta}}.
	\end{equation}
	\subsection{SPH discretized form of governing equation}
	Using the basic discrete equations of the SPH, and considering discontinuous thermal conductivity $k$ we can obtain the SPH discretised form of general heat diffusion equation \eqref{eq:1} \cite{Cleary1999} \cite{Biriukov2019}
	\begin{equation}
		\frac{du_{i}}{dt} = \sum _{j} \frac{m_{j}}{\rho_{i} \rho_{j}} \frac{4 k_{i} k_{j}}{k_{i} + k_{j}} \frac{T_{i} - T_{j}}{r_{ij}} \frac{dW_{ij}}{dr_{ij}}.
	\end{equation}
	By setting $c_{v}$ and $\rho_{i}$ to 1, and taking convective boundary condition into account, the SPH discretised form of equation \eqref{eq:7} is
	\begin{equation}
		\frac{dT_{i}}{dt} = \sum _{j} \frac{4 k_{i} k_{j}}{k_{i} + k_{j}} \frac{T_{ij}}{r_{ij}} V_{j} \frac{dW_{ij}}{dr_{ij}} - \sum _{j} h (T_{\infty} - T_{j}) V_{j} \frac{dW_{ij}}{dr_{ij}}, \label{eq:13}
	\end{equation}
	where particle i has multiple neighbouring particles j, $V_{j} = \frac{m_{j}}{\rho_{j}}$, $T_{ij} = T_{i} - T_{j}$, $r_{ij} = r_{i} - r_{j}$. The equation was implemented in SPHinXsys diffusion algorithm for robin boundary condition of heat transfer.\\
	Equation \eqref{eq:13} illustrates the discretisation of the temperature calculation by considering internal heat conduction and boundary heat convection. For numerical validation, the steady state heat flow over the window frames should be calculated. At steady state, the heat flow from the hot side to the frame is equal to the heat flow from the frame to the cold side. Therefore, we can obtain the absolute heat flow by considering only one side. For convenience, we use the convection term on the boundary to calculate steady state heat flow. The convective boundary condition in equation \eqref{eq:13} denotes the gradient of heat flux $\bigtriangledown q$, unit $W/(m^{3})$, 
	\begin{equation}
		\bigtriangledown q = \sum _{j} h (T_{\infty} - T_{j}) V_{j} \frac{dW_{ij}}{dr_{ij}}.
	\end{equation}
	Then, to get the heat flow Q ,unit $W$, we need to multiply by a volume term of the smoothed particles on the boundary, we get
	\begin{equation}
		Q = \bigtriangledown q \cdot V_{j}
		= \sum _{j} h (T_{\infty} - T_{j}) V_{j}^{2} \frac{dW_{ij}}{dr_{ij}}.
	\end{equation}
	The thermal simulation of window frames in this paper is a two dimensional heat transfer problem. We consider a unit length in the z-direction, and the value of Q calculated in the programme is the heat flow per unit length in the z-direction (a.k.a. heat flow rate), unit $W/m$.
	\subsection{Time integration scheme}
	In order to ensure the numerical stability of the steady state heat transfer simulation, we need to define a suitable time integration scheme for updating temperature field. For the explicit integration of thermal diffusion equation in heat transfer process using SPH, the time step size should fulfil the following conditions \cite{Tang2022}:
	\begin{equation}
		\bigtriangleup t = \frac{0.5 \rho c_{p} h^{2}}{k}, 
	\end{equation}
	where $\rho$ is the density of fluid particle, $c_{p}$ specific heat capacity at constant pressure, $k$ thermal conductivity, and $h$ the smoothing length. For solid heat transfer problems, $c_{p}$ should be replaced by $c_{v}$. The values of $\rho$ and $c_{v}$ are set to 1 for computational convenience, and the equation can be rewritten as
	\begin{equation}
		\bigtriangleup t = \frac{0.5 h^{2}}{k_{max}},
	\end{equation}
	where $k_{max}$ is the maximum thermal conductivity among the window frame materials.\\
	The configurations of particle interaction are updated once per time step, smaller time step determines the time step size of entire thermal state updating. Therefore, we need to find the maximum thermal conductivity during multiple materials definition in order to determine the minimum diffusion time step and ensure the stability of the entire calculation process. This definition of time step size provides a stability criterion for handling the calculation reliability for diffusion problems with anisotropic materials.\\
	In this paper, we focus only on the steady state heat transfer in the window frames defined in ISO10077-2:2012, and therefore do not perform a dynamic analysis of the transient temperature distribution across the frame. The time integration scheme illustrated here describes an iterative step between two adjacent computational operations and is not a true real-time step for transient thermal analysis.
	\section{Thermal conditions of window frames}
	\subsection{Equivalent thermal conductivity of air cavities}
	There are many different types of air cavities within the window frame construction. Depending on their geometry and the way in which they are connected to the internal and external environment, these cavities are basically categorised into three different types. For the simulation of heat transfer process in window frames, the equivalent thermal conductivity of air cavities must be calculated, whereby the calculation is different for each type of air cavity depending on these categorisations.
	\subsubsection{Air cavities without ventilation}
\begin{figure}
		\centering
		\includegraphics[width=0.6\linewidth]{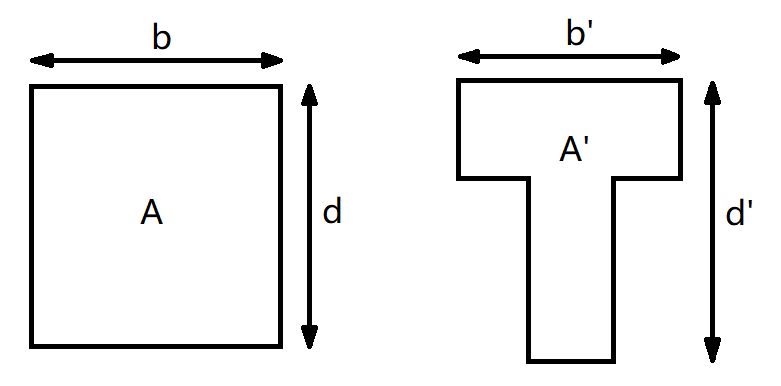}
		\caption[]{Air cavity equivalence}
		\label{fig:ac}
	\end{figure}

	The air cavity is not ventilated at all if its geometry is completely closed or if its geometry is connected to outside of the cavity by a gap of no more than 2 $mm$ in width.
	\paragraph{1. Rectangular geometry}
	For rectangular air cavities without ventilation, equivalent thermal conductivity is defined by
	\begin{equation}
		k_{eq} = \frac{d}{R},
	\end{equation}
	where $d$ is the cavity dimension in the heat flow rate direction, and $R$ is the cavity thermal resistance given by
	\begin{equation}
		R = \frac{1}{h_{a} + h_{r}},
	\end{equation}
	with $h_{a}$ denotes the convective heat transfer coefficient and $h_{r}$ the radiative heat transfer coefficient. These coefficients are defined by
	\begin{equation}
		\begin{cases}
		h_{a} = 
		\begin{cases}
			\frac{C_{1}}{d} & \text{if } b \leq 5mm \\
			max(\frac{C_{1}}{d}, C_{2} \bigtriangleup T^{\frac{1}{3}}) & \text{otherwise}
		\end{cases}\\
		h_{r} = 4 \sigma T_{m}^{3} E F
		\end{cases}.
	\end{equation}
	In our simulation, we consider that $\bigtriangleup T = 10K$ and $T_{m} =283K$, then we get 
	\begin{equation}
		\begin{cases}
		h_{a} = 
		\begin{cases}
			\frac{C_{1}}{d} & \text{if } b \leq 5mm \\
			max(\frac{C_{1}}{d}, C_{3}) & \text{otherwise}
		\end{cases}\\
		h_{r} = C_{4} (1 + \sqrt{1 + (\frac{d}{b})^{2}} - \frac{d}{b})
		\end{cases},
	\end{equation}
	where $C_{3} = 1.57 W/(m^{2}\cdot K)$ and $C_{4} = 2.11 W/(m^{2}\cdot K)$.
	\paragraph{2. Non-rectangular geometry}
	Air cavities with non-rectangular geometry are converted into rectangular cavities with the same area and aspect ratio. The converted rectangular cavity is then used to calculate the equivalent thermal conductivity.
	Figure \ref{fig:ac} shows the equivalent rectangular geometry of a non-rectangular cavity with area $A'$, depth $d'$ and width $b'$.
	The equivalent rectangular cavity of size $b\times d$ and area $A$ must satisfy:
	\begin{equation}
		\begin{cases}
		A = A'\\
		\frac{d}{b} = \frac{d'}{b'}
		\end{cases},
	\end{equation}
	with
	\begin{equation}
		\begin{cases}
		b = \sqrt{A' \frac{b'}{d'}}\\
		d = \sqrt{A' \frac{d'}{b'}}
		\end{cases}.
	\end{equation}
	\subsubsection{Air cavities with slight ventilation}
	Air cavities that are connected to the outside of cavities by a gap of more than 2 $mm$ but not more than 10 $mm$ are considered slightly ventilated.
	The equivalent conductivity is twice as high as that of an unventilated air cavity of the same size. 
	\subsubsection{Air cavities with full ventilation}
	Apart from the two types of air cavities mentioned above, all other cases are considered to be well ventilated air cavities. In this case, it is assumed that the surface is exposed to the environment and should be considered as a developed surface with surface resistance at the boundary.
	\subsection{Boundary condition}
\begin{figure}
		\centering
		\includegraphics[width=0.6\linewidth]{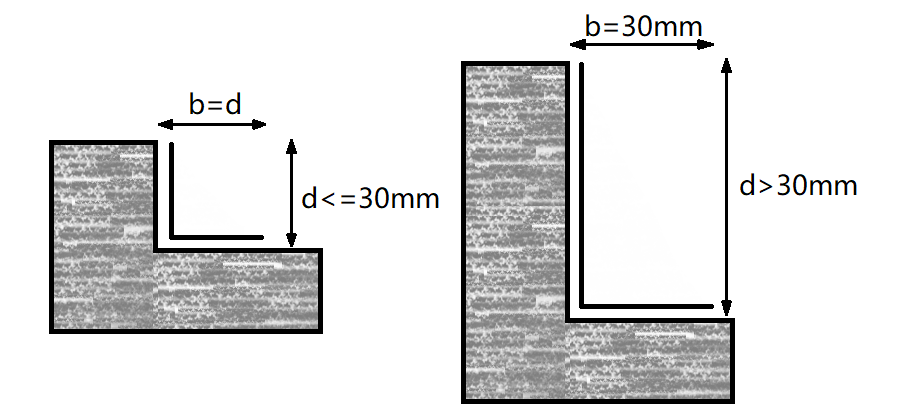}
		\caption[]{Increased surface resistance at corner area}
		\label{fig:isr}
	\end{figure}

	The thermal boundary conditions for window frames in this work consist of two different convections. Internal convection is characterised by high temperatures and high surface resistance, with condensation occurring at the surface transitions. In contrast, external convection is characterised by low temperatures and low surface resistance, which are exposed to normal wind conditions without surface condensation.\\
	The heat flux condition for convections are given by Newton’s law of cooling
	\begin{equation}
		-k\bigtriangledown T = h(T_{\infty} - T_{s}),
	\end{equation}
	where $T_{\infty}$ is the temperature of surroundings, $T_{s}$ the temperature on boundary surfaces, and $h$ the heat transfer coefficient given by
	\begin{equation}
		h = \frac{1}{R},
	\end{equation}
	where $R$ is the surface resistance.\\
	As mentioned above, there are internal and external convections for window frames. According to ISO 10077-2:2012 Annex B \cite{ISO10077-2:2012} and the definition in ISO6946-2017 \cite{ISO6946:2017}, the surface resistance for different convections is shown in Table \ref{tab:sr}.
	\begin{table}
		\scriptsize
		\centering
		\caption{Surface Resistance for Different Profiles}
		\begin{tabularx}{12cm}{@{\extracolsep{\fill}}lcc}
			\hline
			Position & Rse ($m^{2}\cdot K/W$) & Rsi ($m^{2} \cdot K/W$)\\
			\hline
			plane surface & $0.04$	& $0.13$  \\
			\hline
			reduced convection at the junction of two surfaces & $0.04$	& $0.20$  \\
			\hline
		\end{tabularx}
		\label{tab:sr}
	\end{table}
	At the inner boundary, the surface resistance increases due to air condensation at the geometry corner. The location of the increase in surface resistance is shown in Figure \ref{fig:isr}. If d is greater than 30 $mm$, we set b to 30 $mm$. Otherwise, we select b = d \cite{ISO10077-2:2012}.
	In addition, the boundary in contact with the wall(left side of the model) and the ends of the insulation panels(right side of the model) are considered as adiabatic boundaries. We do not need to define the adiabatic boundary condition in SPH method, as there is no heat exchange if there are no contact particles in the neighbourhood of one particle.
	\section{Numerical validation tests}
	Two different window frame configurations were verified in this paper. In each case, the window frame has a hot inner side with a temperature of $20^\circ \text{C}$ and a cold outer side with a temperature of $0^\circ \text{C}$.\\
	Since different window frames have different geometries, we should define size of particles rather than number of particles in order to ensure simulation consistency. General resolution of the heat transfer simulation is set to 0.001 metres. Then, we can obtain the reference smoothing length by the following equation
	\begin{equation}
		h_{ref} = \frac{h_{r} R_{ref}}{R_{r}},
	\end{equation} 
	where $h_{ref}$ is reference smoothing length, $h_{r}$ is the ratio of reference kernel smoothing length to particle spacing, $R_{ref}$ is the reference resolution, and $R_{r}$ is the ratio of system resolution to body resolution. $R_{r}$ and $h_{r}$ are set to 1.3 and 1.0 respectively. A general anisotropic SPH kernel function is used for the simulation, which is a smooth dirac delta function used to quantify the interactions between particles.\\
	To verify the applicability of the calculation procedure to the window frame diffusion problem, it is necessary to calculate the thermal conductivity $L^{2D}$ of the entire section and the thermal transmittance $U_{f}$ of the window frames.\\
	The thermal conductance of entire section $L^{2D}$ is defined by
	\begin{equation}
		L^{2D} = \frac{\phi}{T_{e} - T_{i}},
	\end{equation}
	where $\phi$, unit $W/m$, is the heat flow rate through the window, $T_{e} = 0^\circ\text{C}$ the external temperature and $T_{i} = 20^\circ\text{C}$ the internal temperature.\\
	The thermal transmittance of the frame $U_{f}$, unit $W/(m^{2}\cdot K)$, is defined by
	\begin{equation}
		U_{f} = \frac{L^{2D} - U_{p} b_{p}}{b_{f}},
	\end{equation} 
	where $b_{p}$ is the visible width of the panel expressed in meters, $b_{f}$ the projected width of the frame section expressed in meters, and $U_{p}$, unit $W/(m^{2}\cdot K)$, the thermal transmittance of the central area of the panel.\\
	The calculated thermal conductivity shall not differ from the standard value by more than $\pm 3\%$ and the thermal transmittance shall not differ from the standard value by more than $\pm 5\%$.\\
	The reference values of $U_{p}$, $b_{p}$, $b_{f}$ about different validation cases are shown in Table \ref{tab:rf}, where $U_{p}$ comes from derivations of standard values $U_{f}$ and $L^{2D}$ \cite{ISO10077-2:2012}.
	\begin{table}[htbp]
		\centering
		\begin{tabular}{|c|c|c|c|}
			\hline
			Case & $b_{p}(m)$ & $b_{f}(m)$ & $U_{p}(W/(m^{2}\cdot K))$\\
			\hline
			D2 & 0.19 & 0.11 & 0.551 \\
			\hline
			D4 & 0.19 & 0.11 & 1.034 \\
			\hline
			D7 & 0.19 & 0.048 & 1.169 \\
			\hline
		\end{tabular}
		\caption{Reference value of $U_{p}$, $b_{p}$, $b_{f}$}
		\label{tab:rf}
	\end{table}
	\subsection{Aluminium Clad Wood Frame}
\begin{figure}
	\centering
	\includegraphics[width=1.0\linewidth]{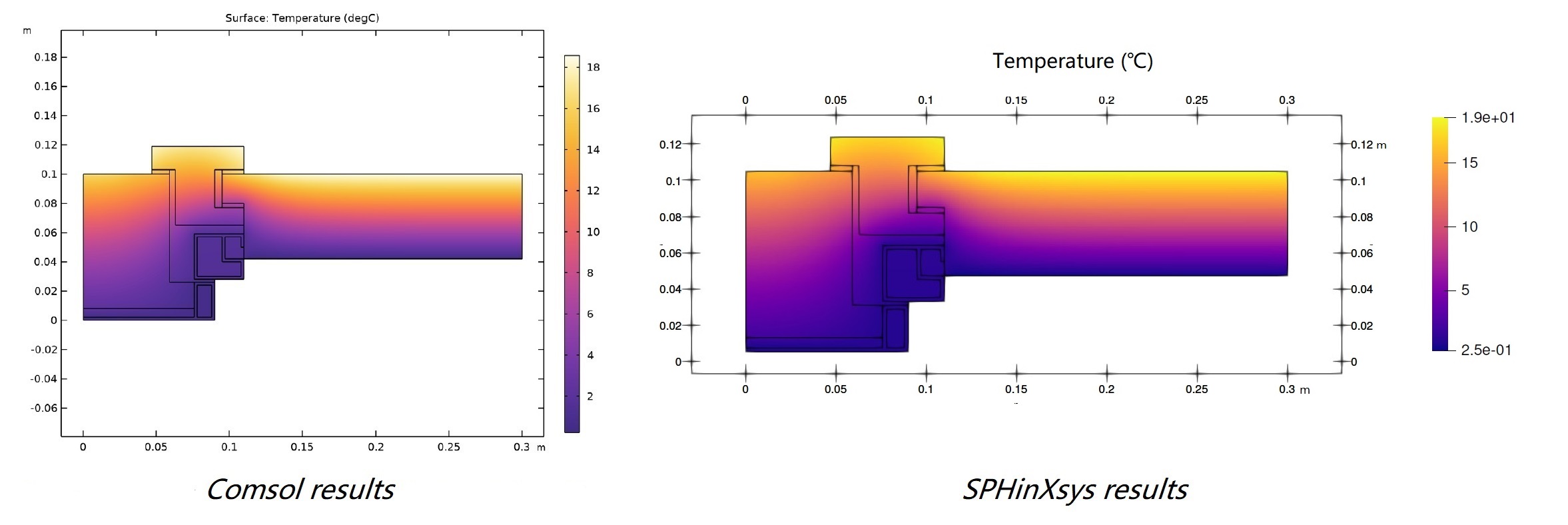}
	\caption[]{Aluminium frame temperature distribution COMSOL vs. SPHinXsys}
	\label{fig:d2}
	\end{figure}

	This application corresponds to case D2 in ISO10077-2:2012. It studies the heat conduction in an aluminium clad wood frame section. The frame is made of two wood blocks with a thermal conductivity of 0.13 $W/(m\cdot K)$. On the external side, a wood block is covered by an aluminium structure which has a high thermal conductivity of 160 $W/(m\cdot K)$. Ethylene propylene diene monomer (EPDM) rubber gaskets are used to waterproof the window. It has a thermal conductivity of 0.25 $W/(m\cdot K)$. The insulation panel has a very low thermal conductivity of 0.035 $W/(m\cdot K)$. Figure \ref{fig:d2} illustrates steady state temperature distribution of the aluminium clad wood frame from COMSOL \cite{Comsol} and SPHinXsys results.
	\subsection{Wood Frame with an Insulation Panel}
\begin{figure}
	\centering
	\includegraphics[width=1.0\linewidth]{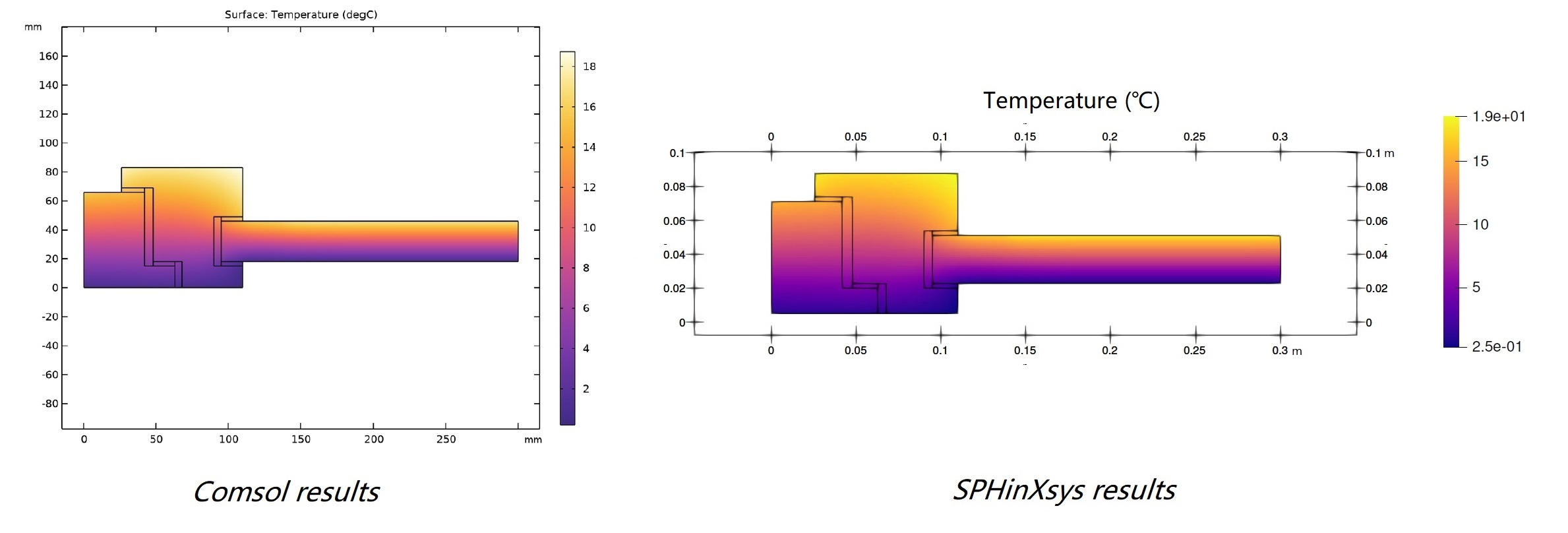}
	\caption[]{Wood frame temperature distribution COMSOL vs. SPHinXsys}
	\label{fig:d4}
	\end{figure}

	This application corresponds to case D4 in ISO10077-2:2012. The frame section is made of two wood blocks with a low thermal conductivity of 0.13 $W/(m\cdot K)$. In order to make the contact between these two blocks and to waterproof the window, two ethylene propylene diene monomer (EPDM) gaskets are used. Two other EPDM blocks are arranged on both sides of the insulation panel. Steady state temperature distribution of the wood frame from COMSOL \cite{Comsol} and SPHinXsys results are shown in Figure \ref{fig:d4}.
	\subsection{PVC Frame with an Insulation Panel}
\begin{figure}
	\centering
	\includegraphics[width=1.0\linewidth]{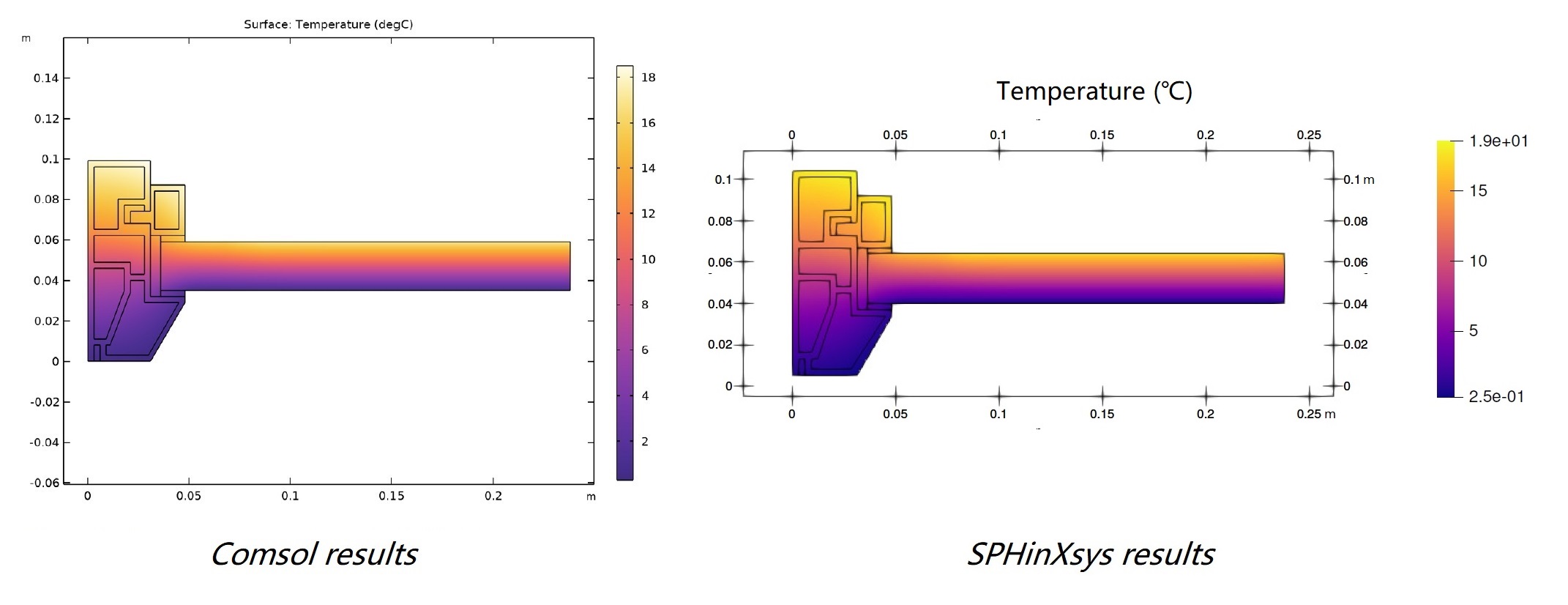}
	\caption[]{PVC frame temperature distribution COMSOL vs. SPHinXsys}
	\label{fig:d7}
	\end{figure}

	This application corresponds to case D7 in ISO10077-2:2012. It studies the heat conduction in a fixed PVC frame section of thermal conductivity 0.17 $W/(m\cdot K)$. Polyamide with a thermal conductivity of 0.25 $W/(m\cdot K)$ is used. There are also some EPDM gaskets to waterproof the window. Figure \ref{fig:d7} demonstrates the steady state temperature distribution of the PVC frame from COMSOL \cite{Comsol} and SPHinXsys results.
	\section{Conclusion}
	The Table \ref{tab:d2v}, Table \ref{tab:d4v} and Table \ref{tab:d7v} present the two validation quantities computed by SPHinXsys alongside the corresponding reference values.
	\begin{table}[htbp]
		\centering
		\begin{tabular}{|c|c|c|c|}
			\hline
			Quantity & Reference value & Computed value & Relative error\\
			\hline
			$L^{2D} (W/(m\cdot K))$ & 0.263 & 0.2629 & -0.04\% \\
			\hline
			$U_{f} (W/(m^{2}\cdot K))$ & 1.44 & 1.4383 & -0.12\% \\
			\hline
		\end{tabular}
		\caption{Validation Quantity of case D2}
		\label{tab:d2v}
	\end{table}
	\begin{table}[htbp]
		\centering
		\begin{tabular}{|c|c|c|c|}
			\hline
			Quantity & Reference value & Computed value & Relative error\\
			\hline
			$L^{2D} (W/(m\cdot K))$ & 0.346 & 0.3408 & -1.50\% \\
			\hline
			$U_{f} (W/(m^{2}\cdot K))$ & 1.36 & 1.3122 & -3.52\% \\
			\hline
		\end{tabular}
		\caption{Validation Quantity of case D4}
		\label{tab:d4v}
	\end{table}
	\begin{table}[htbp]
		\centering
		\begin{tabular}{|c|c|c|c|}
			\hline
			Quantity & Reference value & Computed value & Relative error\\
			\hline
			$L^{2D} (W/(m\cdot K))$ & 0.285 & 0.2834 & -0.56\% \\
			\hline
			$U_{f} (W/(m^{2}\cdot K))$ & 1.31 & 1.2769 & -2.53\% \\
			\hline
		\end{tabular}
		\caption{Validation Quantity of case D7}
		\label{tab:d7v}
	\end{table}
	According to the validation quantity tables, we can clearly see that all three cases satisfy the requirements of the ISO standard. The relative errors for $L^{2D}$ do not exceed $3\%$, and the relative errors for $U_{f}$ do not exceed $5\%$.\\
	This paper demonstrates the reliability of SPH in modelling heat transfer engineering problems related to window frames, and the SPH algorithms developed in SPHinXsys for thermal simulation of window frames has been validated.
	\section*{CRediT authorship contribution statement}
	\textbf{H.T. Ji:} Conceptualization, Methodology, Investigation, Visualization, Validation, Formal analysis, Writing- original draft, Writing- review and editing; \textbf{D. Wu:} Investigation, Writing- review and editing; \textbf{C. Zhang:} Investigation, Writing- review and editing; \textbf{X.Y. Hu:} Supervision, Methodology, Investigation, Writing- review and editing.
	\section*{Declaration of competing interest}
	The authors declare that they have no known competing financial interests or personal relationships that could have appeared to influence the work reported in this paper.
	
	\bibliographystyle{elsarticle-num}
	\bibliography{reference}
	
\end{document}